\documentclass[conference]{IEEEtran}
\usepackage{multicol}
\usepackage{fancyvrb}
\usepackage[utf8]{inputenc}
\usepackage{tabularx}
\usepackage{hyperref}
\usepackage{stackrel}
\usepackage{amssymb}
\usepackage{amsmath}
\usepackage{mathtools}
\usepackage[dvips]{graphics}
\usepackage{graphicx}
\usepackage{psfrag}
\usepackage{subfigure}
\usepackage{dsfont}
\usepackage{stackrel}
\usepackage{theorem}
\usepackage{bigints}
\usepackage{arydshln}
\newtheorem{definition}{Definition}
\newtheorem{theorem}{Theorem}

\newtheorem{lemma}{Lemma}

\newcommand{\defeq}{ \triangleq }
\begin{document}
%\setlength{\abovedisplayskip}{2pt}
%\setlength{\belowdisplayskip}{2pt}
%\setlength{\abovedisplayshortskip}{1pt}
%\setlength{\belowdisplayshortskip}{1pt}
% paper title
\title{Fourier-Motzkin Elimination Software for Information Theoretic Inequalities}
\author{
\IEEEauthorblockN{Ido B. Gattegno}
\IEEEauthorblockA{Ben-Gurion University of the Negev\\
idobenja@post.bgu.ac.il\vspace{-8mm}}
\and
\IEEEauthorblockN{Ziv Goldfeld}
\IEEEauthorblockA{Ben-Gurion University of the Negev\\
gziv@post.bgu.ac.il\vspace{-8mm}}
\and
\IEEEauthorblockN{Haim H. Permuter}
\IEEEauthorblockA{Ben-Gurion University of the Negev\\
haimp@bgu.ac.il\vspace{-8mm}}
}
\maketitle
\IEEEpeerreviewmaketitle
\section{Abstract}\label{abstract}
\par We provide open-source software implemented in MATLAB, that performs Fourier-Motzkin elimination (FME) and removes constraints that are redundant due to Shannon-type inequalities (STIs). The FME is often used in information theoretic contexts to simplify rate regions, e.g., by eliminating auxiliary rates. Occasionally, however, the procedure becomes cumbersome, which makes an error-free hand-written derivation an elusive task. Some computer software have circumvented this difficulty by exploiting an automated FME process. However, the outputs of such software often include constraints that are inactive due to information theoretic properties. By incorporating the notion of STIs (a class of information inequalities provable via a computer program), our algorithm removes such redundant constraints based on non-negativity properties, chain-rules and probability mass function factorization.
This newsletter first illustrates the program's abilities, and then reviews the contribution of STIs to the identification of redundant constraints.
%===================================================================================================================================================
\section{The Software}\label{app_example}
%===================================================================================================================================================
\par The Fourier-Motzkin elimination for information theory (FME-IT) program is implemented in MATLAB and available, with a graphic user interface (GUI), at \url{http://www.ee.bgu.ac.il/~fmeit/}.
The Fourier-Motzkin elimination (FME) procedure \cite{schrijver1998theory} eliminates variables from a linear constraints system to produce an equivalent system that does not contain those variables. The equivalence is in the sense that the solutions of both systems over the remaining variables are the same.
To illustrate the abilities of the FME-IT algorithm, we consider the Han-Kobayashi (HK) inner bound on the capacity region of the interference channel \cite{te1981new} (here we use the formulation from \cite[Theorem 6.4]{el2011network}).
The HK coding scheme insures reliability if certain inequalities that involve the partial rates $R_{10},R_{11},R_{20}$ and $R_{22}$, where
\vspace{-2mm}
\begin{align}\label{example_rjj}
    R_{jj}=R_j-R_{j0},\;j=1,2,
\end{align}
are satisfied.
To simplify the region, the rates $R_{jj}$ are eliminated by inserting (\ref{example_rjj}) into the rate bounds and adding the constraints
\begin{equation}
    R_{j0}\leq R_j,\;j=1,2.
\end{equation}
\begin{figure*}[ht]
    \centering
    \includegraphics[scale=0.72]{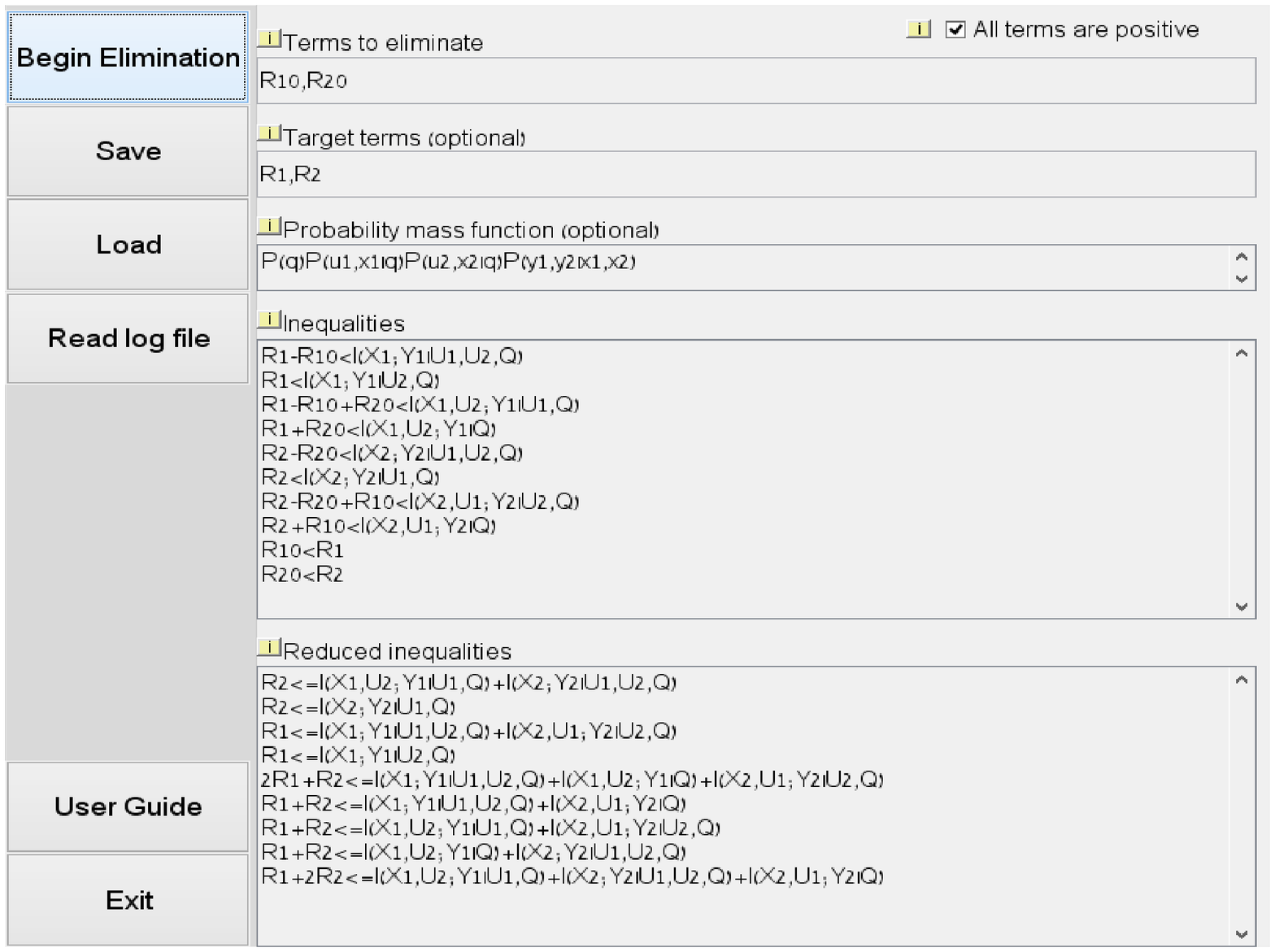}
    \caption{FME-IT input and output - HK inner bound.}
    \label{fig_screen}
\end{figure*}
\vspace{-7mm}
\par The inputs and output of the FME-IT program are illustrated in Fig. \ref{fig_screen}.
The resulting inequalities of the HK coding scheme are fed into the textbox labeled as 'Inequalities'.
The non-negativity of all the terms involved is accounted for by checking the box in the upper-right-hand corner.
The terms designated for elimination and the target terms (that the program isolates in the final output) are also specified.
The joint probability mass function (PMF) is used to extract statistical relations between random variables. The relations are described by means of equalities between entropies.
For instance, in the HK coding scheme, the joint PMF factors as
\vspace{-1mm}
\begin{equation}\label{app_example_PMF}
 \mspace{-1mu}P_{Q,U_1\mspace{-1mu},\mspace{-1mu}U_2\mspace{-1mu},\mspace{-1mu}X_1\mspace{-1mu},\mspace{-1mu}X_2,Y_1\mspace{-1mu},\mspace{-1mu}Y_2}\mspace{-3mu}=\mspace{-3mu}        P_{Q}P_{X_1\mspace{-1mu},\mspace{-1mu}U_1|Q}P_{X_2\mspace{-1mu},\mspace{-1mu}U_2|Q}P_{Y_1\mspace{-1mu},\mspace{-1mu}Y_2|X_1\mspace{-1mu},\mspace{-1mu}X_2},
\end{equation}
and implies that $(X_2,U_2)-Q-(X_1,U_1)$ and $(Y_1,Y_2)-(X_1,X_2)-(Q,U_1,U_2)$ form Markov chains. These relations are captured by the following equalities:
\vspace{-1mm}
\begin{subequations}
\begin{align}
H(X_2,U_2|Q)&=H(X_2,U_2|Q,U_1,X_1)\\
H(Y_1,Y_2|X_1,X_2)&=H(Y_1,Y_2|Q,U_1,U_2,X_1,X_2).
\end{align}
\end{subequations}
%\noindent The output of the program is the simplified system from which redundant inequalities are removed.
%\par Note that although the first and the third inequalities are redundant \cite[Theorem 2]{chong2008han}, they are not captured by the algorithm. This is since their redundancy relies on the HK inner bound
\vspace*{-7mm}
\par The output of the program is the simplified system from which redundant inequalities are removed.
Note that although the first and the third inequalities are redundant \cite[Theorem 2]{chong2008han}, they are not captured by the algorithm. This is since their redundancy relies on the HK inner bound being a union of polytops over a domain of joint PMFs, while the FME-IT program only removes constraints that are redundant for every fixed PMF.
An automation of the FME for information theoretic purposes was previously provided in \cite{joff2008fme}. However,  unlike the FME-IT algorithm, the implementation in \cite{joff2008fme} cannot identify redundancies that are implied by information theoretic properties.
%===================================================================================================================================================
\section{Theoretical Background}
%===================================================================================================================================================
\subsection{Preliminaries}\label{prelimiaries}
%===================================================================================================================================================
\par We use the following notation.
Calligraphic letters denote discrete sets, e.g., $\mathcal{X}$. The empty set is denoted by $\phi$, while $\mathcal{N}_n\defeq\{1,2,\dots,n\}$ is a set of indices.
Lowercase letters, e.g., $x$, represent variables. A column vector of $n$ variables $(x_1,\dots,x_n)^\top$ is denoted by $\mathbf{x}_{\mathcal{N}_n}$, where $\mathbf{x}^\top$ denoted the transpose of $\mathbf{x}$. A substring of $\mathbf{x}_{\mathcal{N}_n}$ is denoted by $\mathbf{x}_\alpha=(x_i\in\Omega\;|\;i\in\alpha,\;\phi\neq\alpha\subseteq\mathcal{N}_n)$, e.g., $\mathbf{x}_{\{1,2\}}=(x_1,x_2)^\top$.
Whenever the dimensions are clear from the context, the subscript is omitted.
Non-italic capital letters, such as $\mathrm{A}$, denote matrices.
Vector inequalities, e.g., $\mathbf{v}\geq\mathbf{0}$, are in the componentwise sense.
Random variables are denoted by uppercase letters, e.g., $X$, and similar conventions apply for random vectors.
%===================================================================================================================================================
\vspace{-1mm}
\subsection{Redundant Inequalities}\label{lp_app}
\par Some of the inequalities generated by the FME may be redundant.
Redundancies may be implied either by other inequalities or by information theoretic properties. To account for the latter, we combine the notion of Shannon-type inequalities (STIs) with a method that identifies redundancies by solving a linear programming (LP) problem.
%===================================================================================================================================================
\subsubsection{Identifying Redundancies via Linear Programming}
%===================================================================================================================================================
\par Let $\mathrm{A}\mathbf{x}\geq\mathbf{b}$ be a system of linear inequalities.
To test whether the $i$-th inequality is redundant,
define
\begin{itemize}
    \item $\mathrm{A}^{(i)}$ - a matrix obtained by removing the $i$-th row of $\mathrm{A}$;
    \item $\mathbf{b}^{(i)}$ - a vector obtained by removing the $i$-th entry of $\mathbf{b}$;
    \item $\mathbf{a}_i^\top$ - the $i$-th row of $\mathrm{A}$;
    \item $b_i$ - the $i$-th entry of $\mathbf{b}$.
\end{itemize}
The following lemma states a sufficient and necessary condition for redundancy.
\vspace{-3mm}
\begin{lemma}[Redundancy identification]\label{lp_app_lemma_identify}
    The $i$-th linear constraint in a system $\rm{A}\mathbf{x}\geq\mathbf{b}$ is redundant if and only if
    \begin{align}\label{lp_app_lemma_identify_eq}
        \rho^\ast_i=\min_{{\substack{\mathbf{x}:\\ \mathrm{A}^{(i)}\mathbf{x}\geq\mathbf{b}^{(i)}}}}\mathbf{a}_i^\top\mathbf{x}
    \end{align}
    satisfies $\rho^\ast_i\geq b_i$.
\end{lemma}

Lemma \ref{lp_app_lemma_identify} lets one determine whether a certain inequality is implied by the remaining inequalities in the system by solving a LP problem. When combined with the notion of STIs, the lemma can also be used to identify redundancies due to information theoretic properties.

%===================================================================================================================================================
\subsubsection{Shannon-Type Inequalities}\label{itip}
%===================================================================================================================================================
\par In \cite{yeung2008information}, Yeung characterized a subset of information inequalities named STIs, that are provable using the ITIP computer program \cite{yeung_itip} (see also \cite{xitip}).
\par Given a random vector $\mathbf{X}_{\mathcal{N}_n}$ that takes values in $\mathcal{X}_1\times\ldots\times\mathcal{X}_n$,
define $\mathbf{h}_\ell\defeq\big(H(\mathbf{X}_\alpha)|\phi\neq\alpha\subseteq\mathcal{N}_n\big)$ \footnote{We order the elements of $\mathbf{h}_\ell$ lexicographically.}.
The entries of $\mathbf{h}_\ell$ are \emph{labels} that correspond to the joint entropies of all substrings of $\mathbf{X}_{\mathcal{N}_n}$.
Every linear combination of Shannon's information measures is uniquely representable as $\mathbf{b}^\top\mathbf{h}_\ell$, where $\mathbf{b}$ is a vector of coefficients. This representation is called the \textit{canonical form}.
Fixing the PMF of $\mathbf{X}_{\mathcal{N}_n}$ to $p$, $\mathbf{h}_\ell(p)\in\mathbb{R}^{2^n-1}$ denotes the evaluation of $\mathbf{h}_\ell$ with respect to $p$.
\par We represent a linear information inequality as $\mathbf{f}^\top\mathbf{h}_\ell\geq0$,
where $\mathbf{f}$ is a vector of coefficients, and say that it \textit{always holds} if it holds for every PMF.
Formally, if
\begin{equation}\label{itip_gamma_star_min}
            \min_{p\in\mathcal{P}}\mathbf{f}^\top\mathbf{h}(p)=0,
\end{equation}
where $\mathcal{P}$ is the set of all PMFs  on $\mathbf{X}_{\mathcal{N}_n}$, then $\mathbf{f}^\top\mathbf{h}_\ell\geq0$ always holds.
\par Since the minimization problem in (\ref{itip_gamma_star_min}) is intractable, Yeung suggested a simple affine space that contains the set where the canonical vectors take values. This space is described by all basic inequalities, which are non-negativity inequalities on all involved entropy and mutual information terms.
The description is further simplified by introducing a minimal set of information inequalities, referred to as \textit{elemental inequalities}.
\vspace{-7mm}
\begin{definition}[Elemental inequalities]\label{itip_const_def_elemental}
    The set of \textit{elemental inequalities} is given by:
    \begin{subequations}\label{itip_const_eq_elemental}
        \begin{align}
        H(X_i|\mathbf{X}_{\mathcal{N}_n\backslash\{ i\}})&\geq0\\
        I(X_i;X_j|\mathbf{X}_{\mathcal{K}})&\geq0,
        \end{align}
        where $i,j\in\mathcal{N}_n,\;i\neq j,\; \mathcal{K}\subseteq\mathcal{N}_n\backslash \{i,j\}$.
    \end{subequations}
\end{definition}

The left-hand side of every elemental inequality is a linear combination of the entries of $\mathbf{h}_\ell$. Therefore, the entire set can be described in matrix form as
\begin{align}\label{itip_const_eq_gamma_n_by_G}
    \mathrm{G}\mathbf{h}_\ell\geq\mathbf{0},
\end{align}
where $\mathrm{G}$ is a matrix whose rows are coefficients.
Consequently, the cone
\begin{align}
\Gamma_n=\Big\{\mathbf{h}\in\mathbb{R}^{2^n-1}\Big|\mathrm{G}\mathbf{h}\geq\mathbf{0}\Big\},
\end{align}
contains the region where $\mathbf{h}_\ell(p)$ take values. The converse, however, does not hold in general.

Based on $\Gamma_n$, one may prove that an information inequality always holds by replacing the convoluted minimization problem from (\ref{itip_gamma_star_min}) with a LP problem. To state this result, we describe the probabilistic relations that stem from the factorization of the underlying PMF by means of linear equalities between entropies (such as in (\ref{app_example_PMF})) as
\begin{align}\label{itip_const_eq_Q}
    \mathrm{Q}\mathbf{h}_\ell=\mathbf{0},
\end{align}
where $\mathrm{Q}$ is a matrix of coefficients.
\vspace{-2mm}
\begin{theorem}[Constrained STIs {{\cite[Theorem 14.4]{yeung2008information}}}]\label{itip_const_theorem_always_holds_gamma_n}
    Let $\mathbf{b}^\top\mathbf{h}_\ell\geq0$ be an information inequality, and let
    \begin{align}\label{itip_const_eq_always_holds_gamma_n}
        \rho^\ast=\min_{\substack{\mathbf{h}:\\ \mathrm{G}\mathbf{h}\geq\mathbf{0}\\ \mathrm{Q}\mathbf{h}=\mathbf{0}}}\mathbf{b}^\top\mathbf{h}.
    \end{align}
    If $\rho^\ast=0$, then $\mathbf{b}^\top\mathbf{h}_\ell\geq0$ holds for all PMFs for which $\mathrm{Q}\mathbf{h}_\ell=\mathbf{0}$, and is called a constrained STI.
\end{theorem}
%===================================================================================================================================================
\section{The Software Algorithm}\label{app_it}
%===================================================================================================================================================

\par The algorithm is executed in three stages. In the first stage, the input system of linear inequalities is transformed into matrix form. Assume the input system contains $L$ variables. Denote by $\mathbf{r}_0$ the $L$-dimensional vector whose entries are the variables of the system. The input inequalities are represented as
\vspace{-6mm}
\begin{align}\label{app_it_AB_matrices}
    \mathrm{A}_0\mathbf{r}_0+\mathrm{B}_0\mathbf{h}_\ell\geq\mathbf{c_0},
\end{align}
where $\mathbf{c}_0$ is a vector of constants and $\mathbf{h}_\ell$ is the vector of joint entropies as defined in Subsection \ref{itip}.
The rows of the matrices $\mathrm{A}_0$ and $\mathrm{B}_0$ hold the coefficients of the rates and the information measures, respectively, in each inequality.
We rewrite (\ref{app_it_AB_matrices}) as
\begin{subequations}\label{app_it_original_matrix_form}
    \begin{align}
        \hspace{-11mm}\mathrm{A}_1\mathbf{x}_1\geq\mathbf{c}_0,
    \end{align}
where
    \begin{align}
        \mathrm{A}_1&\defeq\left[\begin{array}{c|c}\mathrm{A}_0&\mathrm{B}_0\end{array}\right]\\
        \mathbf{x}_1&\defeq(\mathbf{r}_0^{\top} \ \mathbf{h}_\ell^{\top})^\top.\label{EQ:x1_def}
    \end{align}
\end{subequations}
Henceforth, the elements of $\mathbf{h}_\ell$ are also treated as variables.
\par The second stage executes FME. Suppose we aim to eliminate the first $L_0<L$ variables in the original $\mathbf{r}_0$. To do so, we run the FME on the first $L_0$ elements of $\mathbf{x}_1$ (see (\ref{EQ:x1_def})) and obtain the system
\begin{align}
    \mathrm{A}\mathbf{x}\geq\mathbf{c},
\end{align}
where $\mathbf{x}$ is the reduced version of $\mathbf{x}_1$ after the elimination. The matrix $\mathrm{A}$ and the vector $\mathbf{c}$ are determined by the FME procedure.
\par The third stage identifies and removes redundancies. Let
\begin{align}
    \mathrm{\tilde{G}}\defeq
    \left[
        \begin{array}{c|c}
    \mathrm{0}&\mathrm{G}
    \end{array}
    \right],
\end{align}
where $\mathrm{G}$ is the matrix from (\ref{itip_const_eq_gamma_n_by_G}), and
\begin{subequations}
    \begin{align}
    \tilde{\mathrm{A}}&\defeq
    \left[\begin{array}{c}
        \mathrm{A} \\ \hline
        \mathrm{\tilde{G}}
    \end{array}\right] \\
    \tilde{\mathbf{c}}&\defeq\left(\mathbf{c}^\top \ \mathbf{0}^\top\right)^\top.
    \end{align}
\end{subequations}
Further, to account for constraints that are induced by the underlying PMF factorization, set
\begin{align}
    \mathrm{\tilde{Q}}\defeq
    \left[
        \begin{array}{c|c}
    \mathrm{0}&\mathrm{Q}
    \end{array}
    \right],
\end{align}
where $\mathrm{Q}$ is the matrix from (\ref{itip_const_eq_Q}).
Applying Lemma \ref{lp_app_lemma_identify} (redundancy identification) \footnote{We use an extended version of Lemma \ref{lp_app_lemma_identify} that accounts also for equality constraints \cite[Theorem 2.1]{shafiu2007identifying}.} on each of the rows of
\begin{subequations}
\begin{equation}\label{app_it_final_system}
    \mathrm{\tilde{A}}\mathbf{x}\geq\tilde{\mathbf{c}}
\end{equation}
under the constraint
\begin{equation}
\mathrm{\tilde{Q}}\mathbf{x}=\mathbf{0},
\end{equation}
\end{subequations}
while relying on the machinery of Theorem \ref{itip_const_theorem_always_holds_gamma_n}, removes the redundant inequalities and results in the reduced system.
\bibliographystyle{IEEEtran}
\bibliography{IEEEabrv,ref}
\end{document}